# Sensing and control of segmented mirrors with a pyramid wavefront sensor in the presence of spiders

Noah Schwartz*[a], Jean-François Sauvage[b,c], Carlos Correia[c], Cyril Petit[b], Fernando Quiros-Pacheco[d], Thierry Fusco[b,c], Kjetil Dohlen[c], Kacem El Hadi[c], Niranjan Thatte[e], Fraser Clarke[e], Jérome Paufique[f], Joel Vernet[f]

[a]UK Astronomy Technology Centre, Blackford Hill, Edinburgh EH9 3HJ, United Kingdom; [b]ONERA, 29 avenue de la Division Leclerc, 92322 Châtillon, France; [c]Aix Marseille Univ, CNRS, LAM, Laboratoire d'Astrophysique de Marseille, Marseille, France; [d]GMTO Organization, 465 North Halstead St, Suite 250, Pasadena, CA 91107, USA; [e]Dept. of Astrophysics, University of Oxford, Keble Road, Oxford, OX1 3RH, United Kingdom; [f]European Southern Observatory, Karl-Schwarzschild-Str. 2, 85748 Garching, Germany;

## ABSTRACT

The secondary mirror unit of the European Extremely Large Telescope (ELT) is supported by six 50-cm wide spiders, providing the necessary stiffness to the structure while minimising the obstruction of the beam. The deformable quaternary mirror (M4) contains over 5000 actuators on a nearly hexagonal pattern. The reflective surface of M4 itself is composed of a segmented thin shell made of 6 discontinuous petals. This segmentation of the telescope pupil will create areas of phase isolated by the width of the spiders on the wavefront sensor (WFS) detector, breaking the spatial continuity of the wavefront data.

The poor sensitivity of the Pyramid WFS (PWFS) to differential piston (or of any WFS sensitive to the derivative of the wavefront such as the Shack-Hartmann) will lead to badly seen and therefore uncontrollable differential pistons between these areas. In close loop operation, differential pistons between segments will settle around integer values of the average sensing wavelength lambda. The differential pistons typically range from one to tens of time the sensing wavelength and vary rapidly over time, leading to extremely poor performance. In addition, aberrations created by atmospheric turbulence will naturally contain some differential piston between the segments. This differential piston is typically a relatively large multiple of the sensing wavelength, especially for 40 m class telescopes. Trying to directly remove the entire piston contribution over each of the DM segments will undoubtedly lead to poor performance.

In an attempt to reduce the impact of unwanted differential pistons that are injected by the AO correction, we compare three different approaches. A first step is to try to limit ourselves to use only the information measured by the PWFS, in particular by reducing the modulation. We show that using this information sensibly is important but it is only a prerequisite and will not be sufficient. We discuss possible ways of improvement by removing the unwanted differential pistons from the DM commands while still trying to maintain the atmospheric segment-piston contribution by using prior information. A second approach is based on phase closure of the DM commands and assumes the continuity of the correction wavefront over the entire unsegmented pupil. The last approach is based on the pair-wise slaving of edge actuators and shows the best results. We compare the performance of these methods using realistic end-to-end simulations. We find that pair-wise slaving leads to a small increase of the total wavefront error, only adding between 20-45 nm RMS in quadrature for seeing conditions between 0.45"-0.85". Finally, we discuss the possibility of combining the different proposed solutions to increase robustness.

---

* Send correspondence to: noah.schwartz@stfc.ac.uk; phone: +44 (0)131 668 8256

# 1. IINTRODUCTION

HARMONI [1] is a visible and near-infrared integral field spectrograph providing the European Extremely Large Telescope (ELT) with its core spectroscopic capability. It will exploit the ELT's scientific niche in its early years, starting at first light. To fully exploit the spatial resolution and collecting power gain of the ELT, HARMONI will rely on the telescope's adaptive M4 and M5 mirrors. Two different adaptive optics (AO) systems will be used enabling both high-performance combined with low sky coverage using bright natural guide stars (single-conjugate adaptive optics, SCAO) [2, 3] and medium performance combined with excellent sky coverage using 6 laser guide stars and faint natural guide stars for low-order correction (laser tomography adaptive optics, LTAO) [2, 4].

HARMONI will use ELT's deformable quaternary mirror (M4) which contains over 5000 actuators positioned on a nearly hexagonal pattern. The average pitch between actuators is approximately 50 cm when projected onto the primary mirror (M1). M4 is composed of a segmented thin shell made of 6 discontinuous petals (see Figure 1). The segments are discontinuous, but a common silicon carbide (SiC) reference body will ensure that actuators can be driven into position with a known absolute position. In addition, the secondary mirror unit of the ELT is supported by six 50-cm wide spiders, providing the necessary stiffness to the structure while minimising the obstruction of the beam (see Figure 2). Such segmentation of the telescope pupil will create areas of phase isolated by the width of the spiders on the wavefront sensor (WFS) detector, breaking the spatial continuity of the wavefront data. HARMONI will be installed on the Nasmyth platform of the ELT and its SCAO system will use a modulated Pyramid WFS (PWFS).

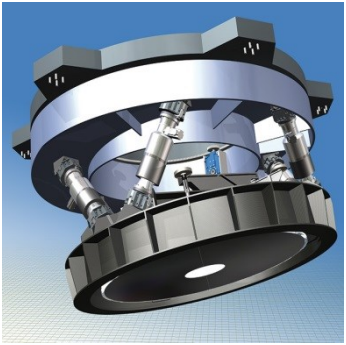
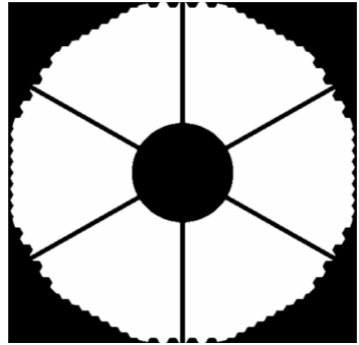

Figure 1: ELT's deformable mirror M4 (credits ESO).     Figure 2: ELT pupil (M1) with the 6 spiders visible.

Previous studies of the impact of pupil fragmentation have mainly been limited to LTAO systems and a simplified 4 petal geometry [5]. They have shown that the impact of the spiders may be smaller when considering small-field tomographic systems such as LTAO. Others have limited their analysis to Shack-Hartmann WFSs for a simplified AO system similar to VLT/SPHERE in size. They have shown that using priors (i.e. minimum variance) has the potential of mitigating the spider effects [6]. Among the possible limitations of these study one can note the continuous deformable mirror surface, the fact that the simulated spiders have a slightly smaller width than what will be available on the ELT or the favourable seeing conditions (close to JQ2, see Table 1).

In this paper, we present realistic numerical simulation results on the performance of the SCAO system for HARMONI in the presence of large spiders, the segmented M4, and a range of turbulence conditions. We show that this configuration can lead to large differential pistons between the DM petals (often termed "island effect") that are injected by the AO loop and we propose simple solutions to mitigate this effect. Numerical simulations are performed using the OOMAO end-to-end modelling tool [7].

The paper is structured as follows. In section 2, we describe the impact of spiders and segmented DMs on SCAO performance. We emphasise the need for small PWFS modulation and careful selection of the valid detector pixels. In section 3 we present results for phase closure, showing the significant limitations of methods ensuring the continuity of the correction wavefront over the entire unsegmented pupil in the presence of large spiders. In section 4 we briefly describe methods to handle differential piston using control algorithms. Section 5 we present a simple method to mitigate differential piston which consist of slaving of edge actuators. Finally in section 6 we conclude this paper and propose potential improvements over the slaving method.

## 2. OF THE IMPACT OF SPIDERS ON SCAO PERFORMANCE

### 2.1 SCAO performance with a continuous DM & without spiders: reference case

The main requirement for the SCAO system of HARMONI is specified for median seeing conditions, a 30° zenith angle, and for a bright natural reference star (magnitude 12 in R). The seeing conditions as per defined by ESO [8] are summarised in Table 1. It shows the equivalent Fried parameter ($r_0$) for the nominal 30° zenith angle, at 60°, and for different wavelengths. The wavefront sensing wavelength for HARMONI is defined in the I-band (i.e. centred on 750 nm). The Fried parameter in median seeing conditions, at this wavelength, and at 30° zenith angle is 23.2 cm. It is important to note that $r_0$ will be smaller than the spiders' width for all seeing conditions, even for the most favourable quartile JQ1. To ensure larger $r_0$, one would for example need to increase the wavelength. In the K-band (i.e. 2.2 µm) the Fried parameter is systematically larger than the spiders' width even for the harshest turbulence conditions JQ4.

Table 1: Seeing conditions and equivalent Fried parameter ($r_0$) in centimetres at 30° zenith angle for different wavelengths. The nominal case for the HARMONI SCAO study is highlighted in bold.

|  | λ | Zenith | Median | JQ1 | JQ2 | JQ3 | JQ4 |
|---|---|---|---|---|---|---|---|
| Seeing [arcsec] | 500 nm | 0° | 0.65 | 0.44 | 0.57 | 0.73 | 1.04 |
| $r_0$ [cm] | 500 nm | 30° | 14.3 | 21.1 | 16.3 | 12.7 | 8.9 |
|  | 750 nm | 30° | **23.2** | 34.3 | 26.5 | 20.7 | 14.5 |
|  | 750 nm | 60° | 10.3 | 15.2 | 11.7 | 9.1 | 6.4 |
|  | 1.6 µm | 30° | 57.6 | 85.1 | 65.7 | 51.3 | 36.0 |
|  | 2.2 µm | 30° | 84.4 | 124.7 | 96.3 | 75.2 | 52.8 |

The WFS for the SCAO system of HARMONI is a 100×100 pyramid WFS (PWFS) running at 500 Hz. We use a classical integrator control with 2 frames delay:

$$u_{n+1} = u_n - g M_{control} s_k$$

Where $M_{control}$ is the control matrix, $g$ is the uniform scalar gain of the integrator, $u_n$ is the DM control vector at time n, and $s_k$ is the slope vector. $M_{control}$ is calculated from the generalised pseudo-inverse of the interaction matrix. Instead of using a zonal approach (i.e. poke matrix) we use a modal approach using Karhunen-Loeve modes (re-orthonormalised over the ELT pupil in Figure 2 and using the actual influence functions of M4). The number of modes we control in the AO loop is selected very coarsly to avoid any overshoot of the controlled modes while still minimising the fitting error. 4760 modes are controlled in weak seeing conditions (i.e. up to 0.85 arcsec) and 4000 modes are controlled above for stronger turbulence. Fine tunning of the number of modes and of the control law is out of the scope of this paper and will be done in a later stage of the project.

In the absence of spiders and using a continuous DM surface we are able to evaluate the SCAO performance in an ideal scenario. It will serve as a reference case and differential piston mitigation solutions will be benchmarked against this reference. Figure 3 shows the average pure SCAO residual error for the reference case. It only takes into account the fitting, servo-lag and noise errors. No other errors (e.g. low-order optimisation loop, M1 co-phasing errors etc.) are taken into account. Using a very simple control strategy we are able to show good overall performance for the studied seeing conditions and very good stability of the residual error at convergence. For example, the pure-AO Strehl ratio in the K-band and at high flux is 92% (101 nm RMS residual error) for a seeing of 0.65 arcsec.

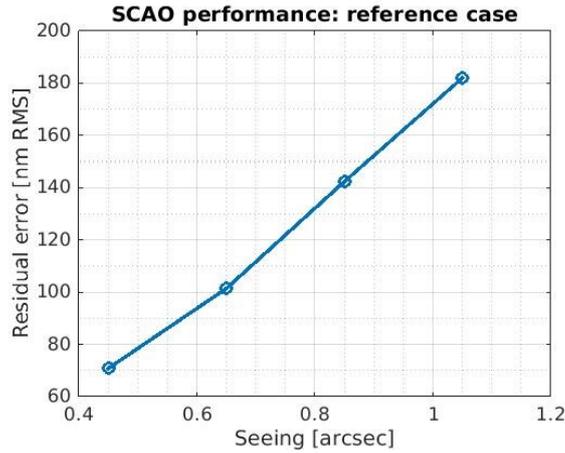

Figure 3: Reference SCAO performance: residual error as a function of seeing condition.

## 2.2 Atmospheric differential piston

Atmospheric turbulence will naturally produce differential piston between each of the six DM segments (often termed petals). Figure 4 shows an example of atmospheric phase in the pupil of the telescope in the presence of spiders (no AO correction). The global piston is set to zero. Figure 5 represents the average piston per segment for the phase presented Figure 4, and shows large natural variations ranging from ±4 μm in this particular example.

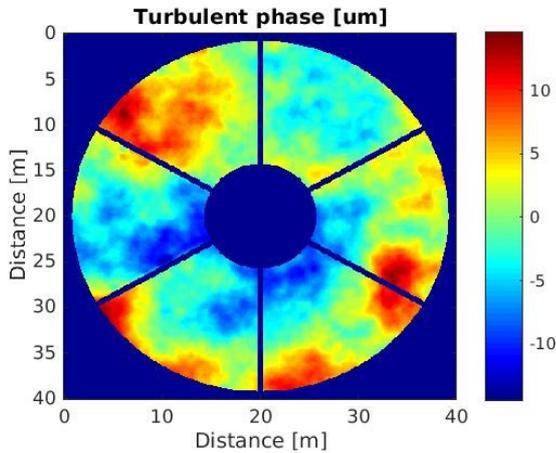
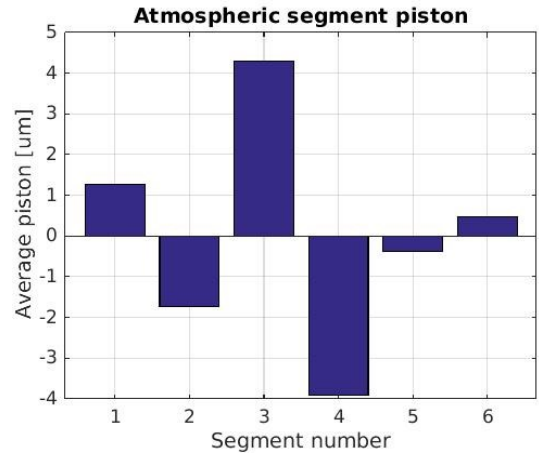

Figure 4: Example of atmospheric phase (in μm) with an average piston over the pupil set to 0. Segments are labelled from the far right and anti-clockwise.

Figure 5: Piston values integrated over each petal segment (in microns) corresponding to the phase in Figure 4. The total piston is set to zero.

Differential piston between segments will evolve as the turbulence changes. Figure 6 shows an example of how the differential pistons of the turbulent atmosphere can evolve over time. It shows that the differential piston can be large (up to ±5 μm in this illustration) and that variations are slow with a typical evolution time over several seconds. The average wind speed in this particular example was set at 5.5 m/s.

Another important aspect to note is that the 50 cm spiders are larger than the coherence length of the atmosphere. This means that the phase on either side of the spiders will be decorrelated. From the phase spatial structure function:

$$D_\varphi(\rho \ll L_0) = \langle |\varphi_r - \varphi_{r+\rho}|^2 \rangle = 6.88 \left( d/r_0 \right)^{5/3}$$

one can expect phase difference to be approximately 1 wave in median seeing conditions (for large distances the structure function will converge to $D_\varphi(\rho \to \infty) = 2\sigma_\varphi^2$). Figure 7 shows the evolution of the phase difference between 2 points in the pupil distant from 50 cm. It varies rapidly over time and can have large value (±1 μm in our example in median seeing conditions).

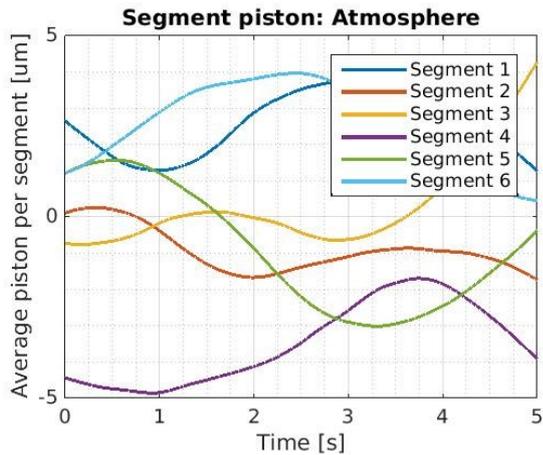 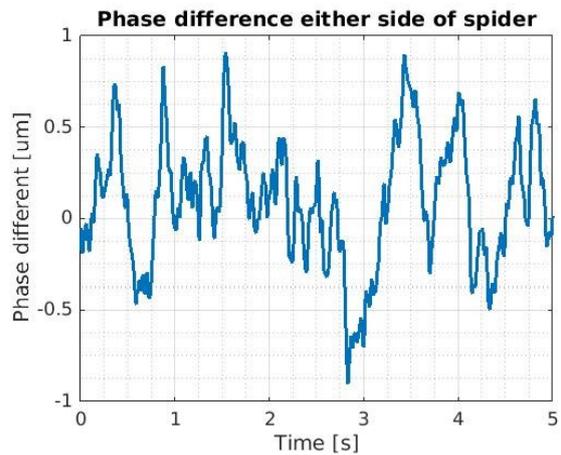

Figure 6: Evolution of atmospheric piston as a function of time. Seeing of 1.05 arcsec at $30^0$ of zenith angle.

Figure 7: Evolution of the different between phase points distant by the spider width (i.e. 50 cm).

### 2.3 Performance in the presence of spiders

If we now add the 6 spiders onto the telescope pupil and take into account the fact the M4 is not a continuous surface but composed of 6 segmented petals, we have a very different outcome from the one obtained in paragraph 2.1. Figure 8 represents the evolution of the residual error as a function of time. It shows a clear divergence of the SCAO loop after a few iterations and a final performance of the order of 5500 nm RMS in median seeing. An illustration of the obtained PSF is given Figure 9.

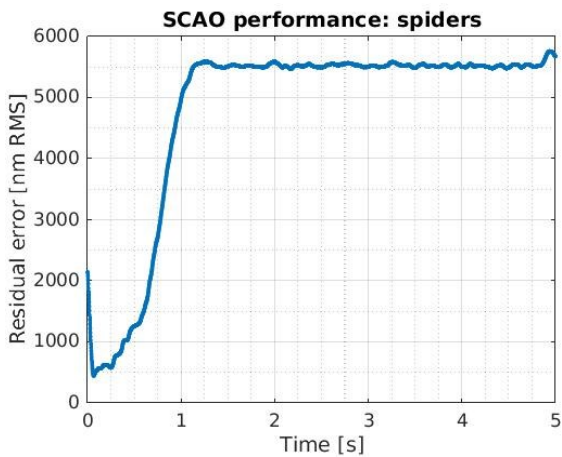 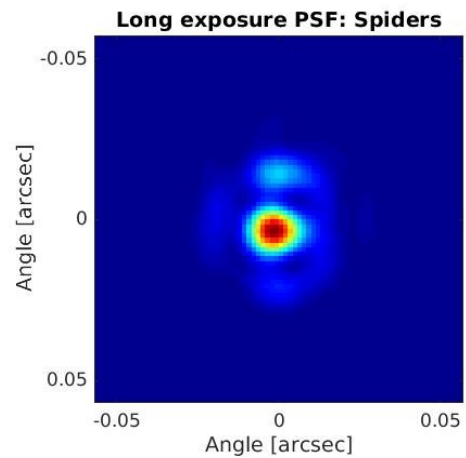

Figure 8: SCAO performance in presence of spider - residual error as a function of time.

Figure 9: Example of PSF obtained in the presence of spiders in median seeing conditions.

The width of the spider is no longer negligible compared to the size of the sub-apertures (the pixel size on the PWFS detector is 37 cm) and will hide full rows of pixels. The spiders divide the measured wavefront slopes map into disconnected domains, with completely missing or corrupted measurements between them. In addition, since the segmented DM is matching the spider geometry, it is fully capable of creating discontinuous modes fitting the 6 petal geometry. Modes that are badly or not sensed at all by the PWFS and that the DM is capable of producing will start to appear. In particular, this is the case for the differential piston as can be seen from Figure 10. Differential piston reaches values up to 6 μm between the petals. In Figure 11 is represented the Eigen modes of the system that are the most badly seen; differential piston can be clearly identified.

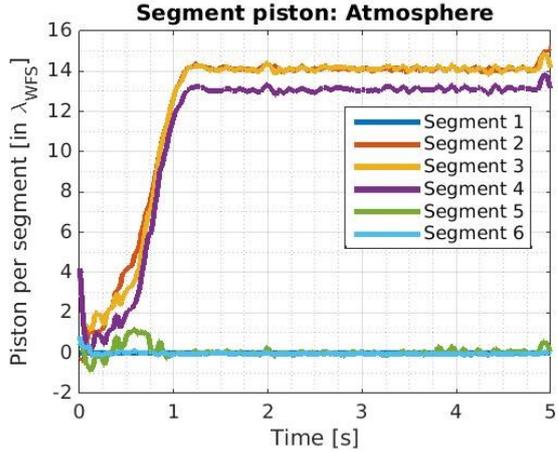 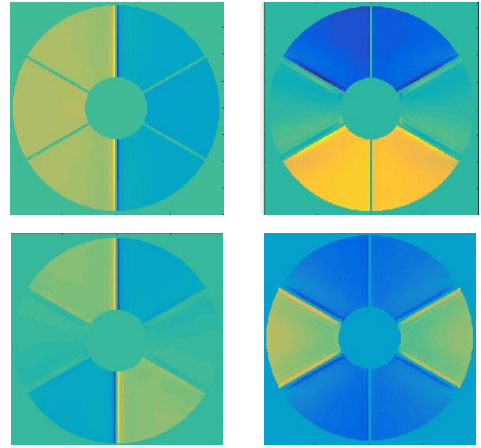

Figure 10: Evolution of the 6 piston values integrated over each petal segment as a function of time. Segment 1 is used as the reference (i.e. piston set to 0). Piston values are scaled by the WFSing wavelength.

Figure 11: The last 4 Eigen modes that have the smaller Eigen values (i.e. the least seen modes).

## 2.4 Differential piston sensing with a pyramid

From Vérinaud et al. [9] we know that the useful PWFS signal that measures differential piston is contained in the diffracted light outside the pupil (i.e. the light that falls in between the pupil segments encodes differential piston). It is therefore important to include the regions under the spiders' shadow when selecting the valid pixel map on the PWFS. In addition, it has been shown that diffracted light outside the pupil footprint comes with small modulation. This means it is important to keep the PWFS modulation as small as possible. Our analyses have shown that a modulation between 3 and 5 $\lambda/D$ is a good choice for the SCAO system of HARMONI. All the simulations presented in the paper have been obtained with a modulation of $3\lambda/D$.

The PWFS signal generated by a differential piston error in the pupil is composed of a phase term that has a sine dependence to an introduced phase step $\Delta$ [10]:

$$S(\Delta, \lambda) \propto sin\left(4\pi\Delta/\lambda\right)$$

This means that there will be an ambiguity in the measurement and in particular for null signals. Even a small residual error (such as a residual tip-tilt for example) may create a large enough differential piston such that the AO system settles at a different piston value for each petals (i.e. modulo $\lambda$). In addition, the signal amplitude will decrease as the length of the gap increases [11], making sensing with large gaps far more challenging. As an example, the signal variance will be down from a normalised value of 1 with a gap of zero length to approximately 0.06 for a 50 cm gap. This cumulative effect creates the island effect shown in section 2.3.

One of the proposed solutions to deal with segmented mirror has been proposed by the Giant Magellan Telescope (GMT) and in part consists in using 2 PWFS sensing at slightly different wavelengths $\lambda$ and $\lambda+\delta\lambda$ to lift the measurement ambiguity [12]. In addition of the extra cost, this solution has the clear disadvantage of increasing the complexity of the system and therefore the associated risks. In the context of HARMONI, we would like to provide a simple and robust solution to handling the unwanted differential piston added by the AO loop which is incompatible with the 2 PWFS solution.

A number of alternative solutions to mitigate the island effect only involving the measurements provided by a single PWFS may also be suggested. For example, differential piston may be attenuated by using a longer sensing wavelength, such that the atmosphere coherent length becomes greater than the spider width and that in parallel atmospheric differential piston also becomes smaller that the sensing wavelength. For the seeing conditions defined in Table 1, this requires increasing the sensing wavelength to the H or K-bands, which in the case of HARMONI is used entirely by the science path. Alternatively, since we know that the information directly under the spiders is not sufficient to completely sense differential piston, adding extra information may improve performance. For example, by defocusing the PWFS it is possible to spread the light under the spiders. Bond et al. [13] have proposed a Fourier-based data extrapolation method that can be applied to extrapolate signals under the spiders. However, since the wavefront on either side of the spiders are decorrelated, adding complementary information cannot be done easily and these solutions exhibit poor performance.

## 3. PHASE CLOSURE

### 3.1 Method

A potential way to mitigate the impact of the island effect is to minimise the differential piston between segments by means of phase closure. Phase closure ensures the continuity of the correction phase going around M4 azimuthally (i.e. 'closure'). Unfortunately, it is not possible to use the piston value measured directly from the average voltages over the DM segment (or actual actuator positions if available). By using this information, the phase closure algorithm will remove the piston component entirely including the term introduced by the atmosphere. We want to correct differential piston introduced by the turbulent atmosphere but still ensure that we only remove the component introduce by the AO loop itself. This is illustrated by Figure 12 and Figure 13 where we can distinguish between the natural differential piston introduced by the atmosphere $\delta_{ATM}$ that we want to correct and the AO introduced differential piston $\delta_{AO}$ that we want to avoid.

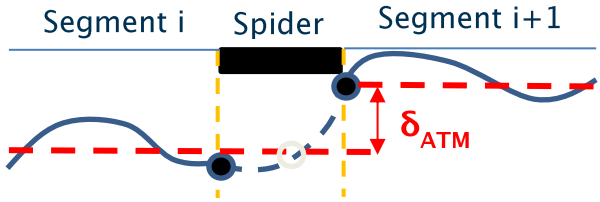
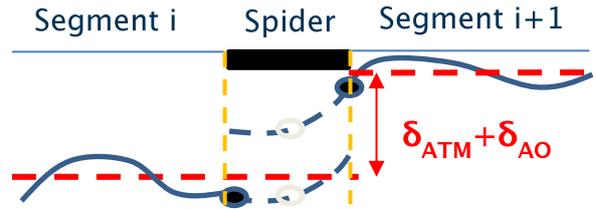

Figure 12: Natural piston difference introduced by the atmosphere between 2 adjacent segments i and i+1. The blue lines represent the wavefront and the dashed red lines the average atmospheric piston.

Figure 13: Differential piston introduced by the AO loop and the atmosphere between 2 adjacent segments. The blue lines represent the wavefront and the dashed red lines the average piston value.

We need to infer the piston to be removed by other means. This information can to some extend be derived from the actuator extensions located at the edge of each segment. In fact, we use the average extension for all actuators directly located alongside the edges (i.e. radial spatial average). We then try to reduce the difference between the phase from one edge of a segment and the phase from the edge of another. Phase closure around M4 is ensured using least square minimisation.

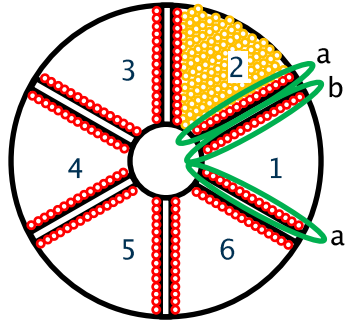

Figure 14: Illustration of the M4 deformable mirror with 6 segments. The actuator locations are depicted by the coloured circles and the red circles represent the edge actuators used to estimate the correction phase at the edge of each segment.

We assume that some information on piston - noted $\alpha_a^i$ and $\alpha_b^i$ - can be extracted for the segment $i$ by using the edges $a$ and $b$. The local piston on segment $i$ is noted $\delta_i$. The M4 DM and the above notation are illustrated by Figure 14. We then assume that the differential at contiguous edges can provide information on differential pistons, such that between segment 1 and 2 we have the following relation:

$$\alpha_b^1 - \alpha_a^2 = \delta_1 - \delta_2$$

All the relations between the other segments follow the same equation and a least square solution can be written as a simple matrix solution. We also set the global piston to 0. From Figure 6 we can see that the atmospheric differential piston has slow dynamics so that it is possible to perform time averaging to improve piston estimates.

### 3.2 Results

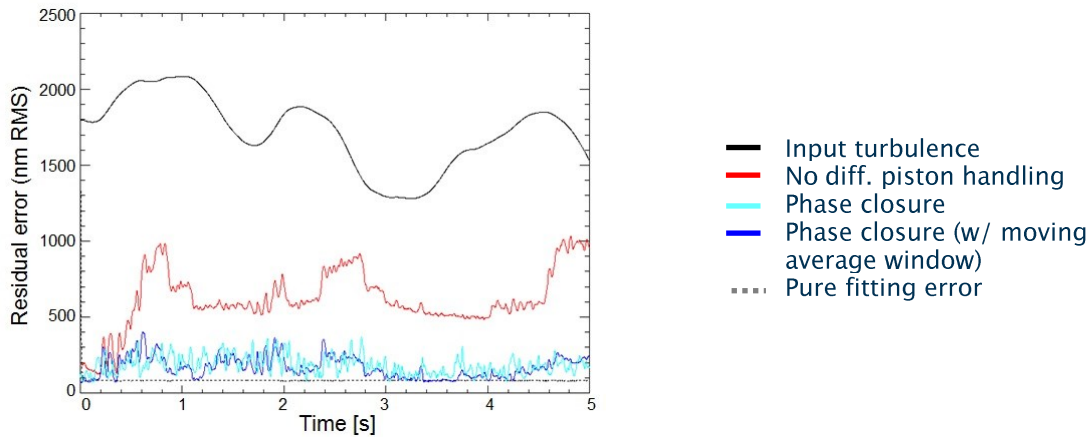

Figure 15: Phase closure: residual error as a function of time for different cases.

The main results are illustrated in Figure 15 and can be summarised as follows:

- A clear improvement in average residual error is observed. From several thousands of nm RMS we are able to reduce the residual error down to an average of 165 nm RMS in nominal seeing conditions (or a Strehl ratio of 86% in the K-band). This value is to be compared to 101 nm RMS obtained in the reference case.
- A large variation of the residual error is still present (minimum 71 nm RMS & maximum 371 nm RMS).
- In an attempt to improve the estimation of the piston values $\alpha_a^i$ and $\alpha_b^i$ we included more rows of actuators along the edges of the segments and performed a linear or spline interpolation. This did not improve the results significantly.
- Using the actuators' extension is slightly less efficient than using the actual phase deformation introduced by the DM actuators.
- Temporal averaging by means of a moving average window slightly improves the results.

In conclusion, this method provides a clear improvement but does not deliver a satisfactory level of correction for HARMONI. We believe this method is ultimately limited by the fact that the gaps are larger than $r_0$. There is a loss of continuity because spiders are larger than the atmospheric coherence length, and the phase on either side of the spider is decorrelated. The natural jump between segments is large and can be as large as several waves. *In fine*, we use a biased estimation of the differential piston to ensure phase continuity and compensate for it.

## 4. HANDLING DIFFERENTIAL PISTON FROM THE COMMANDS

### 4.1 Filtering out segment piston

An obvious method to remove differential pistons between the DM segments would be to ensure that the DM is not capable of producing these modes at all. This can, for example, be achieved by removing the petal piston modes from the Karhunen-Loeve basis to filter out all differential pistons. However, and as discussed previously (see Figure 12 and Figure 13), the turbulent atmosphere itself contains differential pistons that the AO loop needs to compensate.

The correction obtained by such a method is therefore very poor. We use a truncated correction phase to correct for atmospheric turbulence leaving aside large error terms. In addition, as the turbulence strength is increase the atmospheric differential piston term will also increase. In short, this method is clearly not acceptable in term of performance.

### 4.2 Penalisation

An alternative approach would be to add a penalty on the DM commands:

$$u = (M^T M + \alpha V^T V)^{-1} M^T S$$

Where $V$ contains the mode to be rejected such that $v_i^T u < \epsilon$. The parameter α allows for selectivity and trade-off. There are a number of modes that one can penalise. For example, it is possible to penalise the 1$^{st}$ derivatives of the wavefront, curvature of the wavefront or a steep step at the edges of the DM. However, finding a mode that will not be present in atmospheric turbulence but will be produced by the AO loop is very difficult since turbulence itself contains differential pistons. In addition, the trade-off (tuned by the parameter α) is often difficult to make and may change at the loop rate.

### 4.3 Pseudo-open loop control

A promising solution is to use prior information as has been shown by [5, 6] using a regularised MMSE wavefront reconstruction and control algorithms [14]. In particular, prior information on the phase spatial and temporal statistics can be used to smooth the DM commands and will help to keep DM commands free of local pistons. We are currently assessing the performance of such a solution for HARMONI.

## 5. SLAVING EDGE ACTUATORS

### 5.1 Method

We have seen in section 2.1 that performance obtained without any telescope spiders and with a continuous DM surface can be very good. As the spider location and size are fixed by the telescope design there's very little we can change about their thicknesses or any other of their characteristics. M4 actuators are driven into position to a known and absolute distance from the reference body. In addition, all 6 DM segments will use a common SiC reference body and exact actuator extensions across the entire DM surface will be available. It is therefore possible to modify the behaviour of the DM by acting on the extension of the actuators. In particular, we can emulate the behaviour of a DM with a continuous surface. This is done by coupling edge actuators together as illustrated Figure 16 and Figure 17 for 2 actuators located respectively on segment 1 and segment 2. The coupling is done for all pairs of actuators directly opposite of each other.

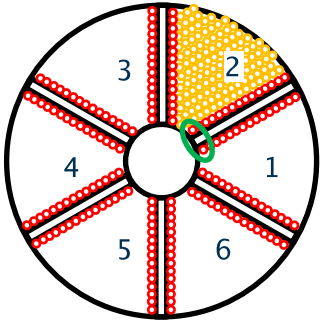

Figure 16: Illustration of M4 with 6 segments. The actuator locations are depicted by the coloured circles and the red circles represent the edge actuators. The coupled actuators seen Figure 17 are circled in green.

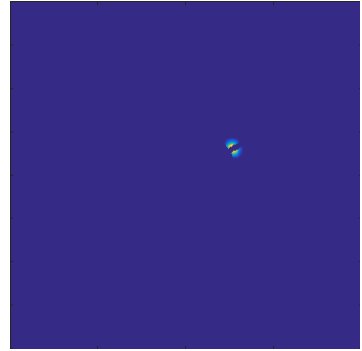

Figure 17: Illustration of the influence created by the coupled actuators highlighted Figure 16. The gap created by the spider is clearly visible.

This method will naturally lead to a reduced number of degrees of freedom, but the reduction in the total fitting error is negligible (e.g. in median seeing conditions, the fitting is increased from 85 nm to 86 nm). For example, the DM will not be able to perfectly correct for an incoming tip-tilt aberration. Figure 18 and Figure 19 illustrate an incoming 800 nm RMS tilt aberration and the residual error after correction by the slaved actuator DM of 2.6 nm RMS. The residual error is mainly located around the spider edges but is clearly negligible.

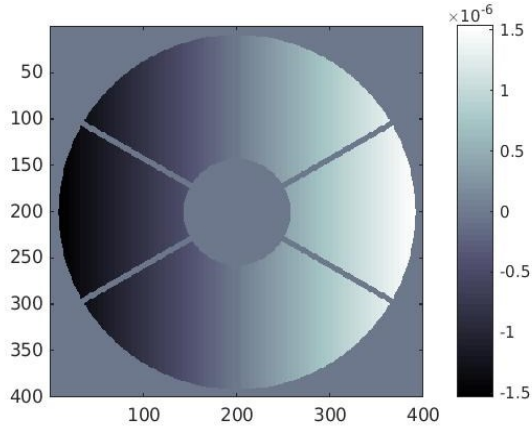

Figure 18: 800 nm RMS tilt aberration.

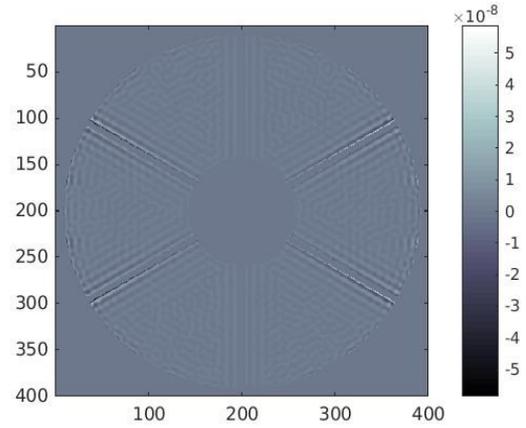

Figure 19: Residual phase after correction of the tilt shown Figure 18 using a DM with coupled actuators. Total residual error: 2.6 nm RMS.

## 5.2 Results

The main results are summarised in Figure 20 where the SCAO performance without spiders and with a continuous DM are presented in red, and SCAO performance with spiders, a segmented DM and the slaving of edge actuators is in blue. The curve in yellow represents the quadratic difference between the slaved actuator DM and the reference case. In other words this represents the additional error term introduced by the islands effect that is not corrected. Our goal, as per defined during the SCAO error budget study, was to ensure that the additional differential piston error was below 70 nm RMS in median seeing conditions (i.e. 0.65 arcsec).

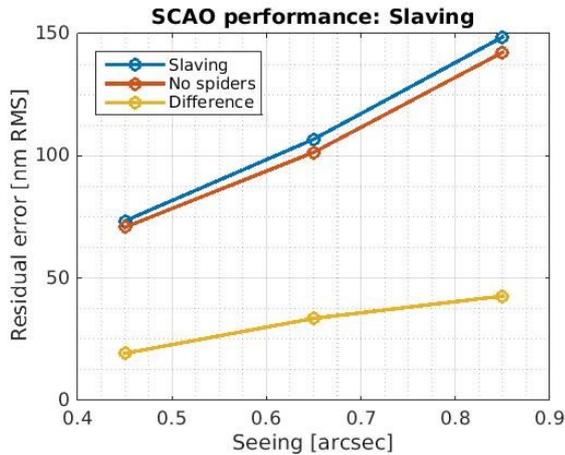

Figure 20: SCAO performance with a slaved DM: residual error as a function of seeing condition.

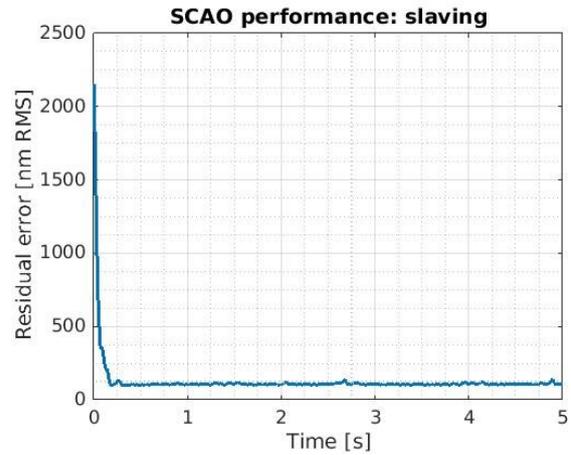

Figure 21: SCAO residual error as a function of time for median seeing condition at 30° zenith angle.

We see that the average residual error in median conditions is 106.8 nm RMS; or an extra 34 nm RMS added to the reference case. This is far below the allocated additional error of 70 nm RMS. Results for weaker (i.e. 0.45 arcsec) and stronger (i.e. 0.85 arcsec) seeing conditions are also presented. Figure 21 shows the AO residual error as a function of time. The convergence of the AO loop will typically be reached after 200 ms (at a loop rate of 500 Hz). Slaving edge actuators removes most of the unwanted differential piston, but a small amount is still present accounting for the small performance variations that can be seen at convergence. The minimum obtained value is 99 nm RMS, while the maximum is 139 nm RMS. The standard deviation at convergence is 5 nm RMS, which is small enough to have a minimal impact on the long-exposure PSFs.

## 6. CONCLUSIONS

In this paper we have studied realistic scenarios for an SCAO system installed at the European Extremely Large Telescope (ELT). The framework of this study is HARMONI, a first-light instrument for the ELT using a 100×100 modulated pyramid wavefront sensor (PWFS) sensing in the I-band. We have shown that large telescope spiders and segmented DMs, dividing the pupil into disconnected domains, can create very large residual errors. The differential piston naturally present in atmospheric turbulence can lead to large phase differences on either side of the spiders. The PWFS is sensitive, albeit poorly, to differential pistons but only modulo π, leading to an ambiguous measurement and phase jumps between DM petals.

We studied and proposed several alternatives to mitigate the island effect; some based on acting on the PWFS-only (e.g. modulation, selection of the valid detector pixels), others acting on the control or the DM itself. The main limitation comes from the large distance between one side of the spider and the other, leading to uncorrelated phases on either side. Methods such as phase closure will use biased information to ensure continuity across the pupil and doesn't deliver the required level of correction.

We propose a simple and robust solution to handle unwanted differential pistons which relies on position/voltage control (i.e. slaving the edge actuators) combined with a small PYR modulation. It also relies on knowing the absolute position of the 6 DM petals and their actuators. This is possible because M4 will use a rigid common reference body to measure the absolute extension of all actuators of the DM. Slaving edge actuators removes most of the unwanted differential pistons, but a small amount is still present. Combining this method with others may improve the performance and stability significantly. We are currently using a single scalar gain for all controlled modes and a possible improvement may be brought by using an optimal modal gain for an optimal control of the filtered modes. In addition, combining with control algorithms using prior information may also improve the behaviour.

Finally, we have demonstrated that a pair-wise of actuators (i.e. slaving) performs very well in SCAO for turbulence conditions up to JQ3 and slightly above. Further analysis is required to demonstrate good performance in strong seeing conditions (JQ4 and above) and in LTAO. In addition, we need to ensure this solution is compatible with the force actuators used for M4 and study the influence of the natural guide magnitude in particular for dim stars.